# Visible light enhanced field effect at $LaAlO_3/SrTiO_3$ interface

Y. Lei[1],Y. Z. Chen[2], Y. W. Xie[3], Y. Li[1], Y. S. Chen, S. H. Wang[1], J. Wang[1], B. G. Shen[1], N. Pryds[2], H. Y. Hwang[3], & J. R. Sun[1*]

[1] Beijing National Laboratory for Condensed Matter & Institute of Physics, Chinese Academy of Sciences, Beijing 100190, Peoples' Republic of China

[2] Department of Energy Conversion and Storage, Technical University of Denmark, Risø Campus, 4000 Roskilde, Denmark

[3] Geballe Laboratory for Advanced Materials and Stanford Institute for Materials &Energy Sciences, Stanford University, Stanford, California 94305, USA

⋆ Author to whom correspondence should be addressed; jrsun@iphy.ac.cn

**Electrical field and light-illumination have been two most widely used stimuli in tuning the conductivity of semiconductor devices. Via capacitive effect electrical field modifies the carrier density of the devices, while light-illumination generates extra carriers by exciting trapped electrons into conduction band[1]. Here, we report on an unexpected light illumination enhanced field effect in a quasi-two-dimensional electron gas (q2DEG) confined at the $LaAlO_3/SrTiO_3$ (LAO/STO) interface which has been the focus of emergent phenomenon exploration[2-14]. We found that light illumination greatly accelerates and amplifies the field effect, driving the field-induced resistance growth which originally lasts for thousands of seconds into an abrupt resistance jump more than two orders of magnitude. Also, the field-induced change in carrier density is much larger than that expected from the capacitive effect, and can even be opposite to the conventional photoelectric effect. This work expands the space for novel effect exploration and multifunctional device design at complex oxide interfaces.**

The q2DEG at the heterointerfaces between complex oxides has received wide attention in recent years because of its implementation for novel physics and prospective applications[2]. The q2DEG confined to the LAO/STO interfaces is a representative system that has been extensively studied[2-14], and exotic properties including two dimensional superconductivity[4], magnetism[6], enhanced Rashba spin-orbital coupling[7], and strong electrical field effect[5,8-14] have been observed. Among these, the field effect is particularly interesting. As already demonstrated, the transport behaviour can be tuned by a perpendicular electrical field across STO or LAO, undergoing a metal-to-insulator transition[10] or a tunable superconducting transition[4,14]. On the other hand, a dramatic modifying of the interfacial conductivity can also be gained by absorbing polar molecular or charges above the LAO layer[8,9].

However, the field effect of complex oxide q2DEG is much more complicated than that of the conventional semiconductor devices. Firstly, significant hysteresis of interfacial conductivity can occur when cycling electrical bias through the STO crystal[10] or scanning a biased tip across the LAO layer, the latter leads to conducting nanowires persisting for days[11,12]. These observations suggest that there exist mobile ionic defects, trapped charges or ferroelectric instabilities in the system, yielding additional freedom in controlling the physical properties of the q2DEG. Secondly, the field effect often exhibits two steps[10,15], where an extremely slow process, usually lasting for thousands of seconds, is comparable to or even stronger than a fast one[15]. The slower rate of field effect indicates the existence of a larger activation barrier which cannot be surpassed by achievable electric field. Here, we report on a dramatic effect produced by combined electrical and optical stimuli for the q2DEGs at both amorphous and crystalline LAO/STO heterointerfaces [a-LAO(12nm)/STO and c-LAO(4uc)/STO, respectively]. We found that photoexcitation dramatically enhanced the ability of the gate field to modulate charge carriers, driving the slow field-induced resistance growth into a great jump beyond the scope of a normal field effect. We ascribe this phenomenon to an oxygen-electromigration-induced interface polar phase whose formation is enhanced by light illumination. The present work demonstrates for the first time the mutual reinforcement of the effects of complementary stimuli on complex oxide interfaces.

Figure 1 shows the resistive responses of a-LAO/STO to electrical and optical stimuli. As schemed in Fig. 1a, a gate voltage, $V_G$, between -100 V and 100 V was applied to the back gate of STO while the a-LAO/STO interface was grounded, and the sheet resistance, $R_S$, was recorded in the presence/absence of a light illumination. As shown in Figs. 1b & 1c, without illumination, the application of a $V_G$=-80 V yields two distinct processes marked respectively by a slight jump and a followed steady increase of $R_S$. The first minor jump is the normal gating effect, stemming from the field-induced charges in the backgate-interface capacitor. The latter process is extremely slow, lasting for more than 2000 s without saturation, and produces a $R_S$ increase much larger than the first jump. This process can be well described by the Curie-von Schweidler law $R_S \propto (t-t_0)^{\alpha}$, which implies a wide distribution of the energy barriers that impede the carrier depletion (see Supplementary materials, Fig. S2)[16].

Remarkably, such a field effect is dramatically modified by light illumination. Aided by a light of 32 mW ($\lambda$=532 nm), as shown by the red curve in Fig. 1b, gate field drives $R_S$ into a sudden jump to a steady state of 200-fold resistance, i.e., the slow process has been dramatically accelerated by light illumination. As demonstrated by Fig. 1d, a light of 32 mW pushes the $R_S(V_G\text{=-100V}, P)/R_S(0,0)$ ratio from ~1.2 up to ~202, amplifying the field effect by ~170 fold. Furthermore, even a $V_G$ as low as -5 V can cause a 17-fold $R_S$ growth (marked by an arrow). This bias is only one-tenth of that usually required to get comparable effect using a backgate without light[10,17]. Enhanced field effect was also observed under positive $V_G$, but it is relatively weak (see Fig. 1b and Figs. S3 & S4 in the Supplementary materials).

Notably, this illumination-induced conductivity change is substantially distinct from the conventional photoelectric effect. As shown in Fig. 1e, without gating field, a light of 32 mW only produces a $R_S$ reduction of ~3.2%, indicating a slight increase in carrier density. However, it causes a giant resistance jump when helped by $V_G$=-80 V, rather than the expected drop.

As shown in Fig. 1f, the illumination enhanced field effect is also observed in c-LAO/STO, suggesting that it is a quite universal phenomenon, independent of the conduction type of the interface (here, it is semiconducting for a-LAO/STO and metallic for c-LAO/STO; refer to Fig. S5 of the Supplementary materials) and the crystal structure of the LAO overlayer (crystalline or amorphous).

To gain a further understanding of this illumination-enhanced field effect, we measured the Hall resistance, $R_{xy}$, and capacitance of a-LAO/STO, $C_{a\text{-}LAO/STO}$. From the linear $R_{xy}$-$H$ relation shown in Fig. 2a, the sheet carrier density, $n_s$, of the initial sample is deduced to be $\sim7\times10^{12}$ cm$^{-2}$. There is no detectable change in the $R_{xy}$-$H$ dependence measured immediately after the application of a $|V_G|$=100 V, indicating that the change in carrier density is tiny. The capacitance data shown in Fig. 2c for $P$=0 indicate a charge density of $\sim3\times10^{11}$ cm$^{-2}$, induced in the charging process of the backgate-interface capacitor (refer to Supplementary materials for the estimation of $\Delta n_S$ based on $C_{a\text{-}LAO/STO}$). In contrast, in a light of $P$ =6 mW (the highest intensity available for our Hall-effect measurement system), a $V_G$=-100 V depresses $n_s$ from $\sim7.0\times10^{12}$ to $\sim1.3\times10^{12}$ cm$^{-2}$. This extraordinarily large $\Delta n_S$ is confirmed by the sudden $C_{a\text{-}LAO/STO}$ drop shown in Fig. 2c for $V_G$< -20 V and $P$=32 mW, which suggests the exhaustion of sheet carriers. It is this carrier exhaustion that causes an uneven distribution of the gating field in STO (Fig. 2d), thus the capacitance drop. A large $\Delta n_S$ ($\sim1.1\times10^{13}$ cm$^{-2}$ for a $V_G$ of -200 V) is also detected in illuminated c-LAO/STO (Fig. S6 in Supplementary materials), it is therefore a general feature of the light-aided gating effect of the LAO/STO interface. This change, $\sim5.7\times10^{12}$ cm$^{-2}$, is well beyond the range of the conventional capacitive effect ($\sim3\times10^{11}$ cm$^{-2}$), strongly suggesting that additional mechanisms are at work under light illumination. For a positive $V_G$=100 V, however, light illumination produces minor effect, and the $\Delta n_S$ only grows to $\sim3.6\times10^{11}$ cm$^{-2}$ in a light of $P$=32 mW. This is consistent with the observed small $R_S$ change, and can be ascribed to the normal field effect.

Further insight can be obtained from the dependence of the field effect on light wavelength, $\lambda$. In Fig. 3a we present the illumination effect obtained under the same power ($P$=32 mW) but different wavelength. The tuned value of $R_S$ drops rapidly as $\lambda$ increases from 532 nm to 850 nm, suggesting that photo-excited processes play a key role in the present illumination-enhanced gating effect. As summarized in Fig. 3b ($V_G$=-40 V), a strong-to-weak crossover of the illumination effect occurs at λ~850 nm (~1.4 eV), suggesting that the photoexcitation of trapped electrons accounts for the present observations. As already shown by previous work[18], oxygen vacancies near the STO surface produced deep level in-gap states that locate ~1.3 eV below the conduction band.

The illumination-enhanced field effect could be explained by a light-modulated lattice polarization in the near interface region of STO, due to the electromigration of oxygen vacancies. There have been a number of reports for the presence of room temperature polarized state in bended STO crystal[19], biaxially strained STO film[20], and LAO/STO superlattices[21]. As reported, oxygen vacancies tend to pile up close to the STO surface[22-24]. They are believed to be the origin of the q2DEG at the a-LAO/STO interface[23,24], and might also contribute to the conduction of the c-LAO/STO interface. A very recent work done by Hanzig *et al.*[25] showed that the migration of these oxygen vacancies ($V_O$s) could induce considerable lattice deformations that favour a polarized phase. The formation process of this phase is very slow, lasting for hours as observe here without light illumination. Obviously, this polarization will contribute an extra tuning to the q2DEG, yielding a slow gating effect. In fact, facilitating carrier modulation by a polar layer has been a mature technique, as Hong *et al.*[26] did for the $Pb(Zr_xTi_{1-x})O_3/La_{0.8}Sr_{0.2}MnO_3$ bilayer system, and the tuned carrier density is exactly equal to $\boldsymbol{P}\cdot\boldsymbol{\sigma}$, where $\boldsymbol{P}$ and $\boldsymbol{\sigma}$ are ferroelectric polarization and area vector, respectively.

Electrical field-induced structure deformation in STO can be directly measured by x-ray diffraction. In Fig. 4a we show the evolution of the (002) peak of STO with gating time. Without illumination, the structure deformation is negligible and only a slight low angle extension of the (002) peak is observed for a gating of -100 V for 7 hours. Since the $V_G$ =-100 V adopted here is only one fifth of that used by Hanzig *et al.*[25], the resulted change is much smaller than that previously reported. However, light illumination strongly enhances the field-induced structure change, as demonstrated by the rapid development of an obvious shoulder on the low angle side of the (002) peak in Fig. 4b. A simple estimation shows that the lattice parameter expands from 3.905 to 3.921 Å after 4 hours' gating. It is a combined effect of gating field and photoexcitation since photon alone has no influence on the structure of STO. According to Hanzig *et al.*, lattice polarization will appear accompanying structure deformation for STO. With this in mind, we believe that light illumination enhances the field effect via accelerating the lattice polarization of STO.

As both theoretically[27,28] and experimentally[29,30] evidenced, the most stable configuration for the oxygen vacancies ($V_O$s) in STO is linear $V_O$ clusters when the concentration of $V_O$s is high, with each $V_O$ trapping one electron. This may the case occurring at the interface of LAO/STO since deep in-gap states, ~1.4 eV, have already been observed. With this in mind, we can present a scenario for the illumination-enhanced field effect. As schematically shown in Fig. 5a, oxygen vacancies in the near interface region in STO maybe mainly in the “+1” valence state with one deeply trapped electron, then is relatively insensitive to electric field[27,29]. As a result, the extra tuning caused by the interfacial polar phase is slow and weak. Light illumination will enhance the response of the oxygen vacancy to applied field by driving their valence state from “+1” to “+2” through exciting trapped electron[28]. In this manner, it accelerates the formation process of the polar phase, in which the motion of oxygen vacancies dominates, thus enhances the field effect. As has been revealed by Fig. 4b and Ref. 25, an electrical field is needed to stabilize the interface polarization induced by the migration of oxygen vacancies. Otherwise the polarization will disappear in several seconds. This is consistent with our observation that $R_S$ quickly drops back when the gate bias is removed (Fig. 1b). When the light is switched off, the excited electrons will fall into the oxygen vacancies again. As a consequence, the lattice deformation becomes weak, and accordingly the polarization effect is weakened. This well explains the dropping back of $R_S$ when light is turned off (Fig. 1e). Probably due to the low flux of the x-ray photons, the photo excited structure change is not significantly induced by the x-rays themselves. In conclusion, our present observations have revealed a unique control of the q2DEG confined at the LAO/STO interface with complementary stimuli of electrical field and light illumination. The principle of multi-stimulus regulation proven here could be extended to a wide variety of complex oxide systems with ferroelectric instabilities.

**Methods**

**Sample fabrication**. The samples a-LAO/STO were prepared by depositing an amorphous LAO layer, ~12 nm in thickness, on $TiO_2$-termined (001)-STO substrates (3×5×0.5 $mm^3$) using the pulsed laser (248 nm) ablation technique. In the deposition process, the substrate was kept at ambient temperature and the oxygen pressure at $10^{-3}$ mbar. The fluence of the laser pulses was 1.5 $Jcm^{-2}$, and the repetition rate is 1 Hz. The target-substrate separation is 4.5 cm. A shadow mask was employed to get the Hall-bar-shaped samples. For comparison, sample c-LAO/STO with a crystalline LAO overlayer (4 unit cells in thickness) was also prepared at a temperature of 800°C and the oxygen pressure of $10^{-5}$ mbar. The fluence of the laser pulses was 0.7 $Jcm^{-2}$, and the repetition rate is 1 Hz. After deposition, the sample was in situ annealed in 200 mbar of $O_2$ at 600°C for one hour, and then cooled to room temperature in the same oxygen pressure. The detailed procedures for sample preparation can be found in Ref. 21 for amorphous overlay and in Ref. 8 for crystalline overlayer. The sample for x-ray diffraction study was prepared by depositing through magnetron sputtering a Ti layer, 30 nm in thickness, above a $TiO_2$-termined (001)-STO substrate.

**Measurements**. Ultrasonic Al wire bonding (20 μm in diameter) was used for electrode connection. Four-probe technique was adopted for resistance measurements. The four welding spots were well aligned, and the separation between neighbouring spots is ~0.4 mm. The formula of $R_S \approx (L/W)R$ was adopted for the convention of four-probe resistance to sheet resistance, where $L$ and $W$ are respectively the long and wide dimensions of the measured plane. Transverse electrical field was applied to STO through an Ag electrode underneath STO, and the LAO/STO interface was grounded. The direction from substrate to interface was defined as positive. The applied current for resistance measurements was 1 μA. Lasers with the wavelengths between 532 nm and 980 nm were used in the present experiments. The spot size of the light is ~0.4 mm in diameter, focusing on the space between two inner Al wires. Under the gate voltage of -100 V, the leakage current was ~0.7 nA without illumination and at most ~7 nA under light illumination (refer to Fig. S1 in Supplementary materials). The crystal structure of STO was measured by a Brüker diffractometer (D8 Advanced) in the presence of a gating field and an illuminating light. The gating field was applied to STO through a top (Ti) and a bottom (Ag) electrode. Capacitance was measured by the Precision Impedance Analyzer (Agilent 4294 A), adopting the a.c. amplitude of 0.5 V and the frequencies of 100 Hz and 5 kHz. The data were recorded after an interval of 60 seconds after the application of $V_G$, and the whole measurement from -40 V to 40 V takes 180 s. All data, except for the $R_S$-$T$ relations, were acquired at ambient temperature.

## Acknowledgements

This work has been supported by the National Basic Research of China (2011CB921801, 2013CB921701) and the National Natural Science Foundation of China (11074285 and 11134007). Y. W. X. and H. Y. H. acknowledge the support from the Department of Energy, Office of Basic Energy Sciences, under Contract No. DE-AC02-76SF00515. J. R. S. thanks Prof. J. W. Cai for his help in preparing the sample for structure analysis.

## Author contributions

J. R. S. conceived and designed the experiments, interpreted, together with Y. Z. C. and Y. W.

X., the experimental results and prepared the manuscript. Y. Lei conducted the experiments. S. H. W. carried out the numerical calculation of capacitance. Y. Z. C. and N. P. provided the amorphous samples and undertook the XPS analysis. Y. W. X. and H. Y. H. provided the crystalline samples. Y. Li and J. W. characterized the sample via AFM. Y. S. C. and Lei Y. and Y. Li performed the experiments for interface polarization. B. G. S. oversaw the project. All authors commented on the manuscript.

## Additional Information

**Competing financial interests:** The authors declared no competing financial interests.

**Supplementary information:** Supplementary information accompanies this paper is available Online or from the author.

## Figure captions

**Fig. 1 | Resistive responses to electrical and optical stimuli of the a-LAO/STO interface**. **a**, A sketch of the experimental setup. **b**, Sheet resistance, $R_S$, recorded with and without a $P$=32 mW light illumination while $V_G$ switches among-80, 0, and +80 V. **c**, Enlarged view of the two-step feature of $R_S$ without light illumination. **d**, Gate dependence of normalized sheet resistance, $R_S(V_G,P)/R_S(0,0)$,recorded at the time of 300 s after the application of $V_G$. **e**, Variation of sheet resistance while the light switches between on and off ($P$=32 mW). Gate voltage is kept at 0 or -80 V during the measurements. For $V_G$=0 the change in $R_S$ has been 50-fold amplified for clarity. The wavelength of the light is 530 nm. All measurements were conducted at room temperature. In all cases, the leakage current (<7 nA) was much lower than the in-plane current applied for resistance measurement, 1 μA (Supplementary materials, Fig. S1).

**Fig. 2 | Hall effect and capacitance measurements. a**, Hall resistance, $R_{xy}$, of a-LAO/STO measured with an in-plane current of 10 μA under different gating/illuminating conditions. Without light illumination the data for $V_G$=-100 V cannot be distinguished from those for $V_G$=0, and therefore are not shown here. **b**, Carrier density and sheet resistance as functions of light power, acquired under a fixed $V_G$ of -100 V. Dashed line is the extrapolated $n_S$-$P$ relation. **c**, Capacitance, $C_{\text{-a-LAO/STO}}$, of a-LAO/STO as a function of gate voltage, measured under the a.c. amplitude of 0.5 V and the frequency of 5 kHz. Labels in the figure denote light power (λ=532). **d**, Schematic diagram for the spatial distribution of gating field in a-LAO/STO when the interface is conductive ($V_G$>0 V) or insulating ($P$=32 mW and $V_G$<-20 V). All the measurements were conducted at room temperature.

**Fig. 3 | Field effect measured in different lights. a**, Sheet resistance of a-LAO/STO corresponding to field switching between on and off, collected at a constant light power (32 mW) but different wavelength. For clarity, only the data for $P$=0 and λ=532 nm are shown for $V_G$ =+40V. **b**, Sheet resistance as a function of light wavelength, collected at the time of 200 s for $V_G$<0 and 1000 s for $V_G$>0. All the measurements were conducted at room temperature.

**Fig. 4 | Photoexcitation acceleration of the field-induced structure deformation of STO. a**, X-ray diffraction spectra of the (002) peak of the gated STO, measured through a 30-nm-thick Ti anode. Gate field produces minor effect without light illumination, and only a slight left expansion of the (002) is detected after a gating of -100 V for 7 hours. **b,** Light illumination strongly enhances the field-induced structure change. An obvious shoulder of the (002) peak, which marks the lattice expansion in the near interface region in STO, develops in several minutes of illumination and vanishes as soon as the gate bias and light are removed. The curve for "$V_G$ & laser off" was collected right after the removal of $V_G$ and $P$. Labels

besides the curves indicate the gating time before the θ-2θ scanning. The total time required for each θ-2θ scanning is ~10 minutes.

**Fig. 5 | Schematic diagrams for the migration of oxygen vacancies under electrical field and light illumination**. **a**, The deeply trapped "+1" valence oxygen vacancies, $V^{1+}$, distribute near the interface. The free electrons, some of which contribute to the conductivity of the q2DEG, are adjacent to $V^{1+}$ due to electrostatic attraction. The $V^{1+}$ vacancies have a trend to move along electrical field, *E*. But this process is slow and, as a result, the $V^{1+}$ migration-induced polarized effect is weak. **b**, The $V^{1+}$ vacancies become $V^{2+}$ after the trapped electrons are photo-excited, and thus have a stronger response to *E*, quickly drifting away from the interface. Accordingly, free electrons also move with $V^{2+}$ due to electrostatic attraction. The migration of the $V^{2+}$ produces a downwards polarized layer, and the electron density in the q2DEG is further tuned by this layer. Dark green marks the structure-deformed and lattice-polarized phase.

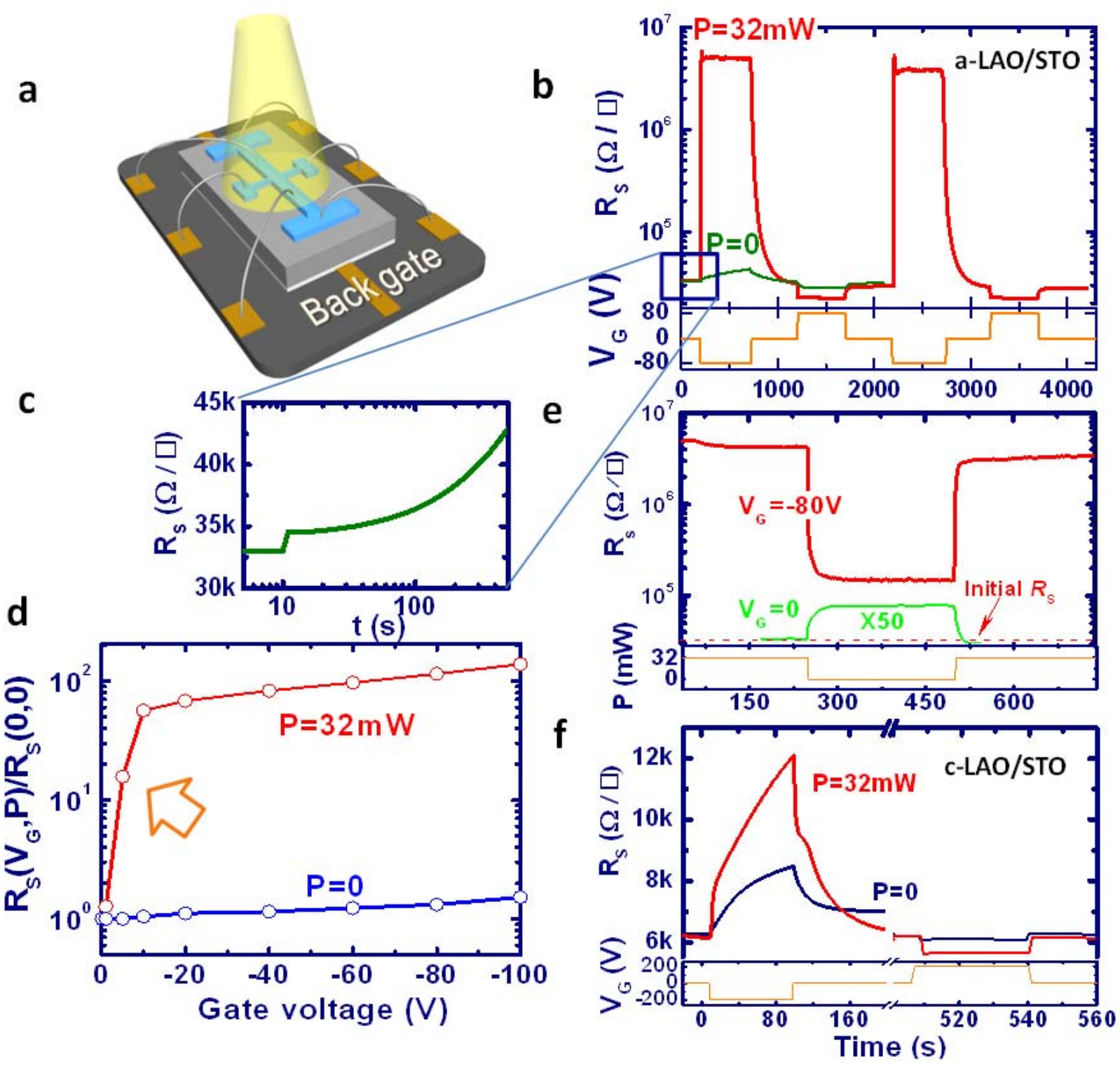

a
b
c
d
e
f
Back gate
P=32mW
a-LAO/STO
P=0
$R_S$ ($\Omega$ / □)
$V_G$ (V)
45k
40k
35k
30k
t (s)
$R_S(V_G,P)/R_S(0,0)$
P=32mW
P=0
Gate voltage (V)
$V_G$=-80V
$V_G$=0
X50
Initial $R_S$
P (mW)
c-LAO/STO
Time (s)

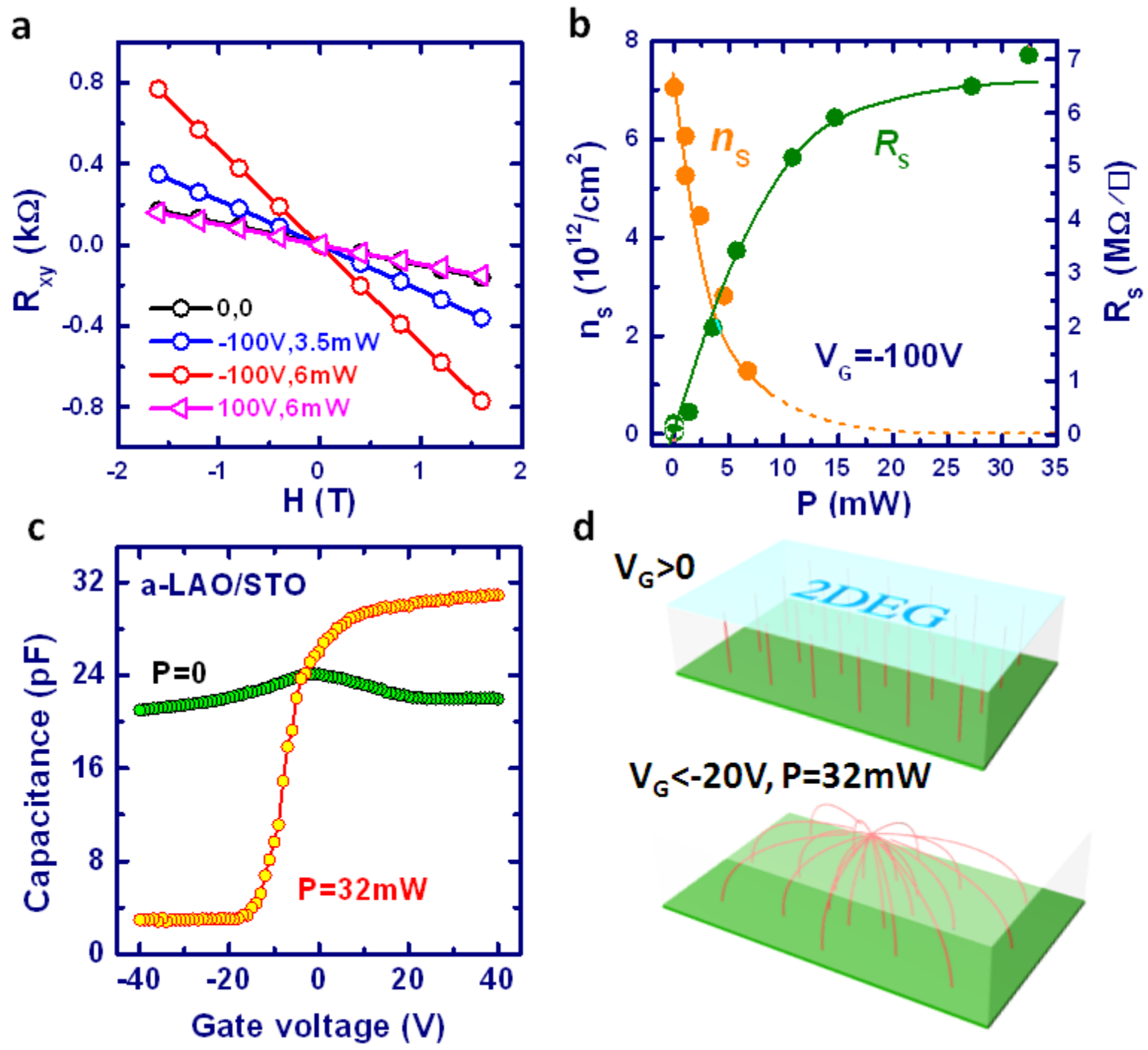
a
0.8
0.4
0.0
-0.4
-0.8
R$_{xy}$ (kΩ)
0,0
-100V,3.5mW
-100V,6mW
100V,6mW
-2
-1
0
1
2
H (T)
b
n$_S$ (10$^{12}$/cm$^2$)
R$_S$ (MΩ/□)
n$_S$
R$_S$
V$_G$=-100V
P (mW)
c
Capacitance (pF)
a-LAO/STO
P=0
P=32mW
Gate voltage (V)
d
V$_G$>0
2DEG
V$_G$<-20V, P=32mW

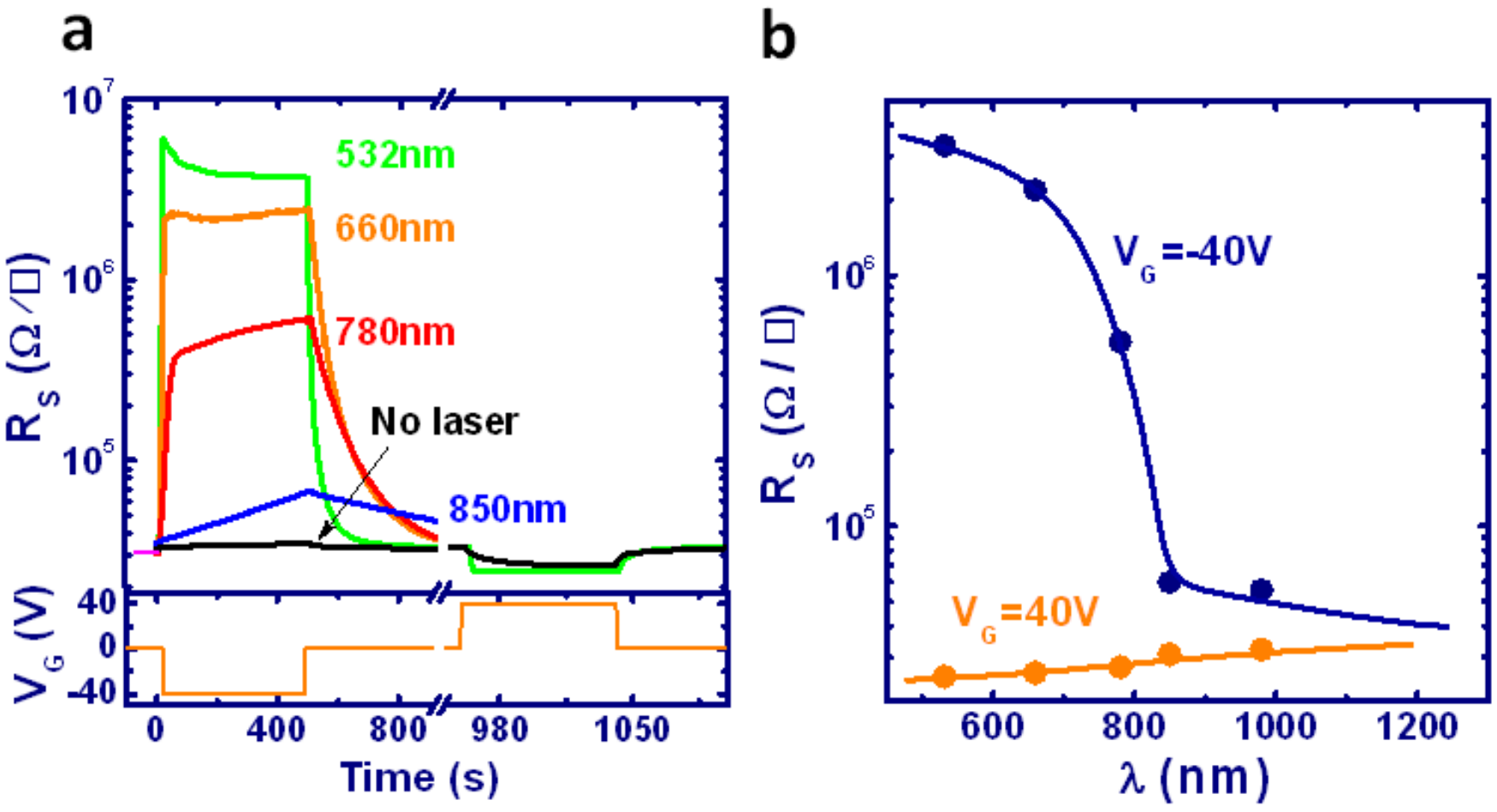

a
b
$10^7$
$10^6$
$10^5$
$R_S$ (Ω/□)
532nm
660nm
780nm
No laser
850nm
$V_G$ (V)
40
0
-40
0
400
800
980
1050
Time (s)
$V_G$=-40V
$V_G$=40V
600
800
1000
1200
λ (nm)

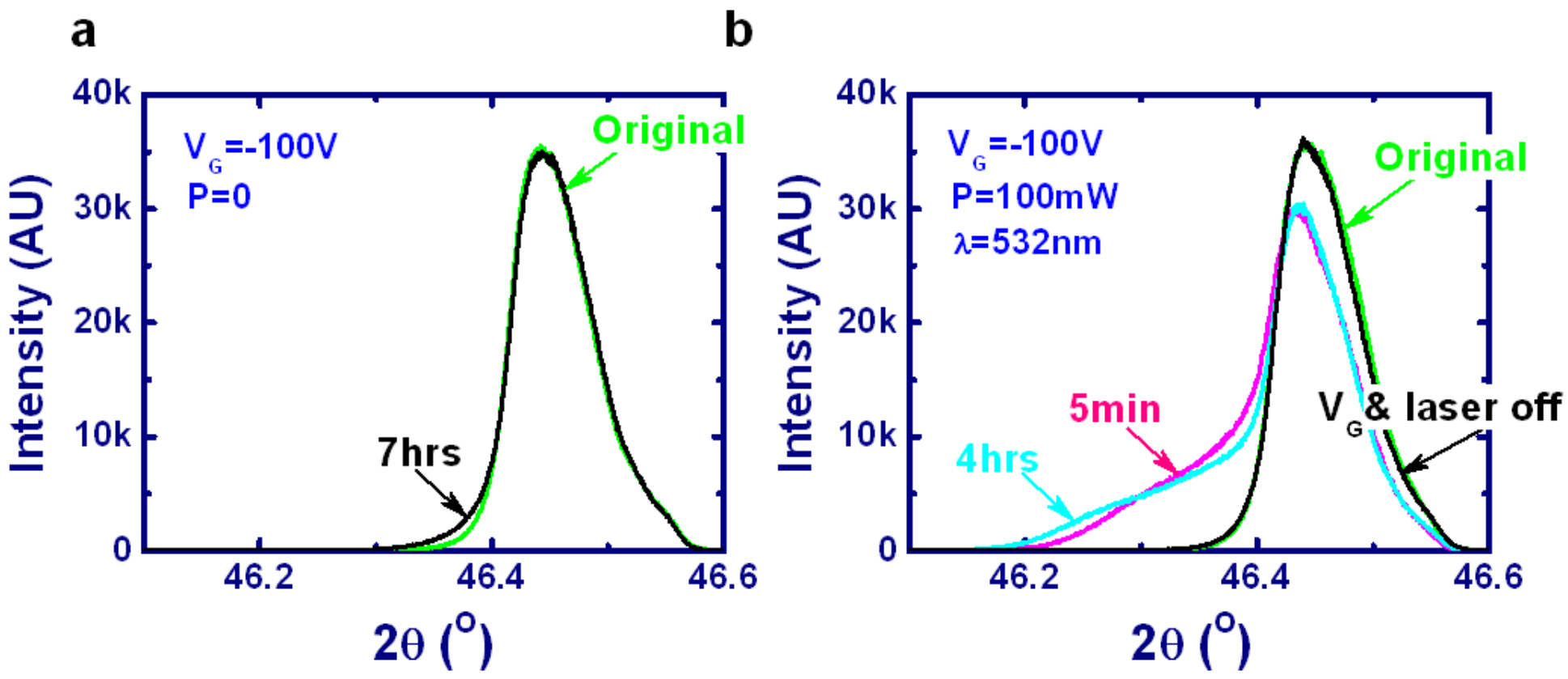
a
b
40k
30k
20k
10k
0
Intensity (AU)
$V_G$=-100V
P=0
Original
7hrs
46.2
46.4
46.6
2θ (°)
$V_G$=-100V
P=100mW
λ=532nm
Original
5min
4hrs
$V_G$ & laser off

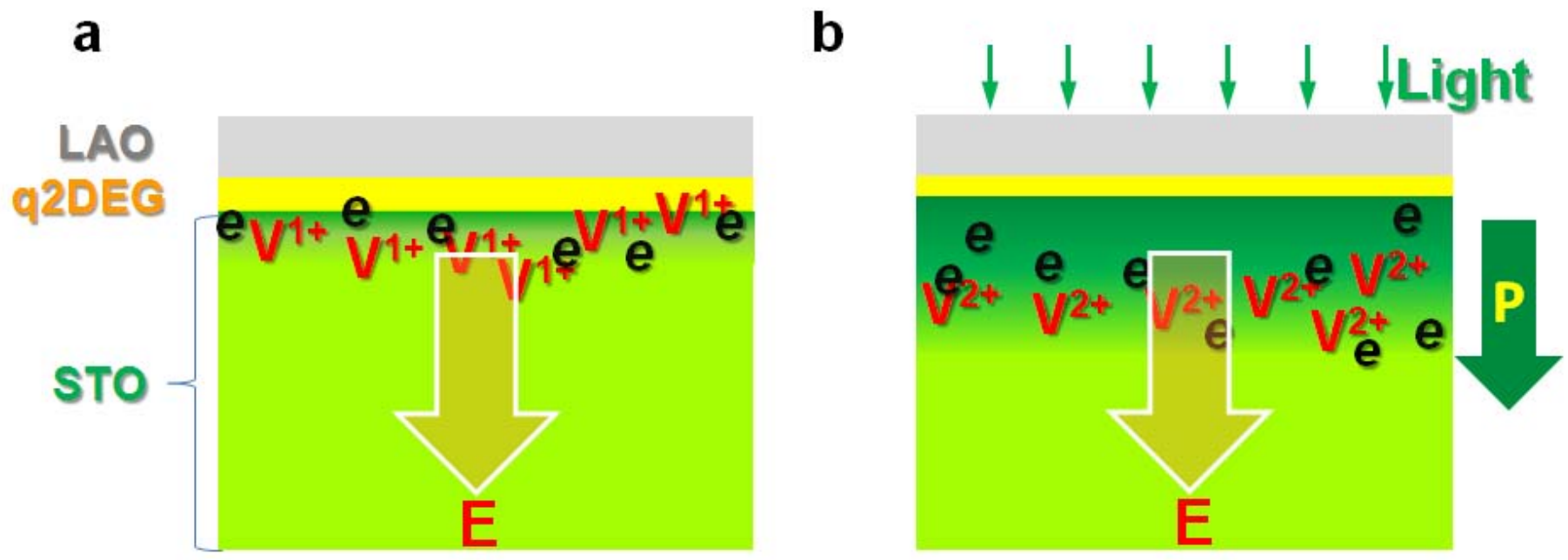
a
LAO
q2DEG
STO
E
b
Light
P
E